\documentclass[prd,aps,reprint,superscriptaddress,floatfix,flushbottom,nobibnotes,nofootinbib,showpacs,amsmath]{revtex4-1}
\usepackage{amssymb}
\usepackage{graphicx}
\usepackage{longtable}
\usepackage{color}
\usepackage{float}

\def\PANDA{$\overline{\mbox{P}}${ANDA}}

\begin{document}
\title{Studies of Hadron Structure and Interactions with the $\overline{{\bf P}}$ANDA Experiment at FAIR}

%\author{  \PANDA{} Collaboration and } \\
%\author{  \begin{center}{\PANDA{} Collaboration and }\end{center}
%\vspace{-7mm} }
%\newline}

\author{ \PANDA{} Collaboration}
\noaffiliation

%\author{Diego~Bettoni}
%\affiliation{Istituto Nazionale di Fisica Nucleare   Sezione di Ferrara, 44122 Ferrara, Italy}

\author{Nora~Brambilla}
\affiliation{Technische Universit\"at M\"unchen, 85747 Garching, Germany}

%\author{Paola~Gianotti}
%\affiliation{Laboratori Nazionali di Frascati, 00044 Frascati, Italy}

%\author{Frank~Maas}
%\affiliation{Helmholtz-Institut Mainz, 55099 Mainz, Germany}

\author{Ulf-G.~Mei{\ss}ner}
\affiliation{Helmholtz-Institut f\"ur Strahlen- und Kernphysik, Universit\"at Bonn, 53115 Bonn, Germany}
\affiliation{Forschungszentrum J\"ulich, 52428 J\"ulich, Germany}

%\author{James~Ritman}
%\affiliation{Forschungszentrum J\"ulich, 52428 J\"ulich, Germany}
%\affiliation{Ruhr-Universit\"at-Bochum, 44780 Bochum, Germany}

%\author{Andreas~Sch\"{a}fer}  
%\affiliation{Institut f\"ur Theoretische Physik, Universit\"at Regensburg,              93040 Regensburg, Germany}  

%\author{Ulrich~Wiedner}
%\affiliation{Ruhr-Universit\"at-Bochum, 44780 Bochum, Germany}

%\collaboration{for the \PANDA{} Collaboration}

\date{\today}
\begin{abstract}

The \PANDA{} experiment together with the high quality antiproton
beam at HESR will be a powerful tool to address fundamental questions of
hadron physics in the charm and multi-strange hadron sector. 
In connection with the
recent data in the hidden charm sector, \PANDA{} will be able to deliver decisive contributions to this field, due to the complementarity of the $\overline{p} p$ entrance channel and the capabilities of the combined storage ring and detector system under construction.

\end{abstract}
%\pacs{12.38.Gc,13.85.-t,14.20.Dh}
\maketitle

\section{Introduction}

%The modern theory of strong interactions, Quantum
%Chromodynamics (QCD), is precisely tested for observables for which 
%perturbative QCD converges well. However, QCD also displays a very rich array 
%of non-perturbative phenomena, many of which are not yet understood.
The modern theory of strong interactions, Quantum Chromodynamics (QCD), displays a very rich
array of non-perturbative phenomena, many of which are not yet understood. 
Progress in this respect will impact a broad range of physical problems,
in settings ranging from astrophysics, cosmology
to strongly coupled complex systems in particle and condensed
matter physics~\cite{QWG1}.
Furthermore, hadronic effects often represent the dominant uncertainties in the tests of various fundamental
quantities in the Standard Model (e.g. the muon $(g-2)$).

\PANDA{} will measure annihilation reactions of antiprotons
with nucleons and nuclei in order to provide complementary
and in part uniquely decisive information on
a wide range of QCD aspects. The $\overline{p} p$ entrance channel
not only couples very strongly
to the gluonic components of produced hadrons and of
allowing resonant production of e.g. quarkonium states, but also allows
a wide range of quantum numbers of the produced system.
Quarkonia, i.e. bound systems of two heavy quarks, have
been instrumental in the establishment of QCD, and
have again a central role, thanks to challenging
and exciting data from BaBar, Belle, BESIII and the LHC.

The scientific scope of \PANDA{} is ordered into several
pillars: hadron spectroscopy, properties of hadrons in
matter, nucleon structure and hypernuclei. Each of these
pillars addresses specific open issues of QCD.

Hadron spectroscopy is currently characterized
by the fact that
many predicted states have not yet been 
found while others were observed, which
do not fit to expectations. 
Candidates for hybrid, molecular, and tetraquark systems
have been discovered. Charged, and thus manifestly exotic
states have been detected.
All these results indicate that
some crucial pieces of the puzzle are missing.
 Success in achieving a theory
description of quarkonium processes and of the new
states impacts our understanding of QCD and strongly
coupled systems at a deeper level and also helps us to
address other fundamental questions in particle physics
and nearby fields~\cite{QWG2}.

\PANDA {} will be able to carry out a comprehensive study of charmonium 
physics, both in direct formation and in production modes (see below), 
providing unique information, which will be complementary  to $e^+e^-$ 
experiments  (Belle2, BESIII, limited to low spin states (mainly vector and 
axial)) and to the LHC, which will study quarkonium production for much 
higher energies. A schematic overview of the relevant states is in 
Fig. 1.
As indicated, there will not only be studies of charm\-onium(-like) states, but also searches for and investigations of states with strong gluonic contributions to the wavefunction, i.e. glueballs and hybrids, are foreseen. 
\begin{figure}[t]
	\begin{center}
		\includegraphics[width=0.26\textwidth]{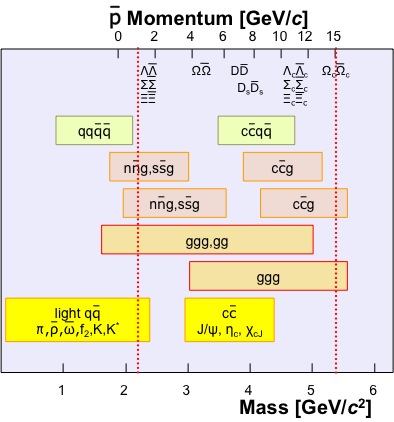}
	\caption{Overview of the spectroscopic states accessible in $\overline{p} p$ annihilation as a function of antiproton momentum (upper scale) or cm energy (lower scale). The red dashed lines indicate the range accessible with \PANDA{} at HESR. }
	\end{center}
\end{figure}

\PANDA{} will address two important issues
connected to hadrons in a nuclear environment.
First, in-medium modifications of hadronic properties 
will give precious information on the mechanism of
chiral symmetry breaking and its partial restoration. 
Second, the measurement
of charm production cross sections
in $\overline{p}$ annihilation on a series of nuclear targets, will allow
to deduce the hidden/open-charm nucleus dissociation
cross sections. These are fundamental to
understand charmonium suppression in relativistic heavy
ion collisions, which is interpreted as a signal for quark-gluon
plasma.

In the study of nucleon structure, recent progress in
theory has concerned the formulation of methods to de-
fine and calculate with nonperturbative approaches parton
physics and generalized distribution functions. Experimental
extractions of these quantities are plagued by
many uncertainties that limit our knowledge of other
quantities. \PANDA{} will be able to contribute in this
field by measuring the crossed-channel counterparts of the
processes that are studied by lepton scattering experiments
in many laboratories e.g. DESY, CERN, JLAB.

For the baryonic sector, \PANDA{} can access many systems for 
detailed studies of spectroscopy and production cross section.
Furthermore, the study of hyperon's production will give
access to hyperon-nucleus and hyperon-hyperon interaction
mechanisms which have implications in particle and
nuclear physics.

To illustrate where \PANDA{} will make  major contributions
in this context, we focus in the following on
just three specific topics for which we discuss not only the
physics, but also in which respects PANDA will be superior
and/or complementary to other experiments. These
topics are 
open charm states, in particular the $D_{s0}(2317$), 
quarkonium 
{and quarkonium-like states, in particular the X(3872), 
and the
field of hypernuclei and hyperons. This small selection
from the many aspects of the \PANDA{} physics program
is made to stress both the importance of charm
physics and the possibility to explore
novel aspects of cold nuclear matter.
%\vspace{-6mm} 

\section{Opportunities with Antiproton Annihilation}

Antiproton-proton annihilation often proceeds via two or three-gluon 
processes. Consequently, as shown at LEAR, this system is very favorable to 
study gluonic degrees of freedom compared to other processes where valence 
quarks and antiquarks can suppress the production of glueballs. 
It was demonstrated by 
the LEAR experiments at CERN \cite{Amsler,Landau} and 
the E760/E835 experiments at Fermilab
\cite{Garzoglio,Bettoni}
that the combination of an 
intense, nearly mono-energetic antiproton beam with a state-of-the-art 4$\pi$ 
experiment is ideally suited
for the field of spectroscopy.
In these experiments not only the most 
prominent glueball candidate, the f$_0$(1500), was discovered, but also 
the relevant PDG results are
dominated by their precision. This is due 
to the unique features offered by cooled antiproton beams combined with 
dedicated detectors. \PANDA{} will be the 
first experiment capable to measure both charged and neutral particles with a 
truly 4$\pi$-detector exploring the energy regime of charm with this high 
precision. 

The use of antiproton-proton annihilation enables two modes to investigate resonances at \PANDA.
In the {\em formation} mode a single resonance is formed directly in the annihilation process, which correspondingly must have $J^{PC}$ quantum numbers 
accessible by a fermion-antifermion pair. In contrast, the {\em production} 
mode involves at least one additional particle to the resonance of interest, 
which thus does not have the same restrictions on $J^{PC}$. The comparison of 
both methods helps to classify the resonances and identify those 
with “exotic quantum numbers”, i.e. quantum numbers forbidden for ordinary 
quark-antiquark mesons. In general, no restrictions for quantum numbers of 
resonances produced in \PANDA{} exist, all states can be populated. Since 
charmonia often decay to final states containing lepton pairs, which are otherwise rare at these 
energies in the hadronic environment, \PANDA{} can often profit from clean 
signal/background ratios.

For those newly discovered X, Y, Z states that possess a very narrow width, a precise determination of the excitation curve is necessary to distinguish between the different theoretical interpretations. This can be done decisively better with \PANDA{} in formation mode due to the strong phase space cooling of the antiproton beam in HESR, compared to the production of these states in the decay chain of heavier particles.

Again, the LEAR experiments demonstrated that the antiproton annihilation is a nearly ideal way to produce baryon-antibaryon pairs with strangeness. This is especially the case close to the corresponding production thresholds, where very clean experimental conditions occur due to the large cross sections and the effective detection of forward-going particles in the forward detector.
In contrast to other experiments, \PANDA{} has a special design consisting of a dipole magnet in the forward spectrometer that bends the accelerator beam away from the zero-degree region. Thus, 
basically the full solid angle for decay products is covered, providing an enormous advantage for the threshold experiments and for the determination of the quantum numbers of states with the help of amplitude analysis. For the production of hypernuclei or for the implantation of strange or charmed baryons in nuclear matter, the existence of a pair of particle and antiparticle provides the unique advantage that the detection of either one of them provides an excellent trigger to select reactions of the partner inside nuclear matter.
%\vspace{-6mm} 
 
\section{Selected Examples}

%\subsection*{The nature of $D_{s0}^*(2317)$}

\noindent {\bf Threshold Scan of the $D_{s0}^*(2317)$}

An excited hadron state for which so far there exists great uncertainty is
the  $D_{s0}^*(2317)$. Can it be described more 
adequately as a two meson molecule with a leading four quark Fock state or as 
a conventional meson with a leading two quark Fock state?
The problem lies in the fact that in QCD one always faces a configuration 
mixture of all field configurations with the same quantum numbers. 
Consequently, this is rather a quantitative than a qualitative question which 
should be formulated as: With which amplitude 
do the leading Fock state components enter the  
$D_{s0}^*(2317)^\pm$ wave function within a specific factorization/renormalization scheme 
at a specific scale? From the theory side, first steps to answer such questions rigorously  
were taken using lattice QCD \cite{LQCD_1,LQCD_2}. While these calculations are still only exploratory they
demonstrate clearly that by the time \PANDA{} will start to take data, precise predictions from the lattice should be available. 
Technically, this requires full-fledged variational calculations with physical quark masses, which will become possible based on the foreseeable increase in computer power within just a few years. 

Experimentally one can test these calculations by measuring precisely the partial widths for the various decay channels of 
the $D_{s0}^{*}(2317)^\pm$ and the other $D_s$ mesons. 
While hadron masses are usually not very specific for the physical nature of the state under discussion, the individual decay widths are. 
For example, the $D_{s1}^{*}(2536)$ decays into $DK$ channels and is thus especially sensitive to any $DK$ component in its wave function,
for instance 
theoretical predictions 
for $D_{s0}^{*}(2317)^\pm\rightarrow D_s+\pi$  range from
$\Gamma = (6\pm 2)$ to 140~keV.    

Once the nature of the $D_{s0}^{*}(2317)^\pm$ is unambiguously clarified by such a combined experimental and lattice approach, 
one can then use the results to formulate a controlled low energy description 
in terms of hadronic degrees of freedom. This description would allow to 
address also those experimental results which are not accessible to LQCD, for 
example real time reaction properties of the $D_s$ states, as illustrated in, 
e.g. \cite{ChPT}.  

The \PANDA{} experiment will be able to make decisive contributions to understanding the nature of the  $D_{s0}^*(2317)$ by performing a high precision measurement of the total width. This is achieved by analyzing the excitation function of the reaction
$\bar p p \rightarrow D_{s0}^*(2317)^\pm D_s^\mp$ within $\pm 2$~MeV of the nominal threshold \cite{PAN09}. Since theoretical calculations of the cross section for this process vary significantly, it is not possible to quote the exact precision for the width determination. For a 1~nb production cross section at 5~MeV over the nominal threshold, the precision to measure the total width varies from 25 to 50~keV, depending upon the actual value of the width. This precision is about two orders of magnitude more sensitive than the current upper limit of 3.8~MeV, which is not expected to be significantly improved below the MeV range by other experiments in the foreseeable future. Thus \PANDA{} will confirm or exclude a large fraction of the predicted range $\Gamma \sim 6-140$~keV expected from theoretical calculations, thereby making a decisive contribution to the understanding of the nature of this potentially exotic state.

%
%\vspace{2mm}

%\subsection{X(3872)}
\noindent {\bf Resonance Scan of the X(3872)}

The $X(3872)$ was first observed by Belle \cite{BelleX} and was subsequently confirmed at BaBar, CDF, D0, LHCb and BESIII. Its main features are a very small width ($\Gamma < 1.2$~MeV), a mass which is close to the $D^0D^{0*}$ threshold and quantum numbers $J^{PC} = 1^{++}$.
As the first new state to be discovered the $X(3872)$ is also the best studied but, despite this, its nature remains unknown. Possible interpretations include conventional charmonium, diquark anti-diquark bound state, a tetraquark state or a molecular state, and the line shape of this state is expected to be a sensitive method to discriminate between the binding mechanisms~\cite{ErikBraaten}.
\par
Theoretical estimates of the cross section  for $X(3872)$ formation in $\bar pp$ annihilations  with a subsequent decay to $J/\psi \pi^+ \pi^-$ have been done in a model-dependent way in~\cite{ChenMa} and range from 2 to 443 nb, with a strong dependence on the total width of this state, for which the authors have assumed the interval between 136 keV and 2.3 MeV.
Experimentally only upper limits are available for the total width ($<$ 1.2\ MeV) and the branching ratio to $\bar pp$ ($< 2 \times 10^{-3}$) and a lower limit for the branching ratio to $J/\psi \pi^+ \pi^-$ ($>$ 0.026) \cite{pdg}.
\par
A precise determination of the width of this very narrow state is difficult in a production experiment ($e^+e^-$ or $pp$), because in this environment the state has to be reconstructed and the achievable precision strongly depends on the experimental detector resolution. On the other hand, a high precision measurement is possible in \PANDA{} by means of a resonance scan in the direct formation process $\bar pp \to X(3872) \to J/\psi \pi^+ \pi^-$.  The availability of $\bar p$ beams with a beam momentum spread $\Delta p/p = 4 \times 10^{-5}$  (high-resolution mode) makes it possible to measure widths of 100\ keV or smaller and makes \PANDA{} a unique experiment to measure these very narrow states. 
This has been proven by means of detailed Monte Carlo studies \cite{Galuska}, in which a 20 point energy scan around the $X(3872)$ was simulated. In order to test the sensitivity of \PANDA{} the simulation was carried out under rather challenging assumptions for the (unknown) resonance parameters: a total width of 100\ keV with a cross section of 50\ nb for the process $\bar pp \to X(3872) \to J/\psi \pi^+ \pi^-$, a rather pessimistic scenario, since the theoretical predictions combine small widths with larger cross sections (or vice versa). This combination of parameters corresponds to $BR(X(3872) \to \bar pp) \times BR(X(3872) \to J/\psi \pi^+ \pi^-) = 3.9 \times 10^{-6}$, perfectly consistent with the current experimental limits. 
The hadronic background was studied using a Dual Parton Model (DPM) based generator~\cite{DPM}.
The results show that the $X(3872)$ can be reconstructed with a signal-to-background ratio of $7$ and its width measured as ($87 \pm 17$)\ keV, consistent with the input width of 100\ keV.
\par
These studies demonstrate that \PANDA{} is unparalleled in its ability to perform resonance scan investigations of such narrow states, which can be formed directly in $\bar pp$ annihilations.

%\vspace{2mm}

%\subsection{Production and Spectroscopy of Baryon-Antibaryon Pairs}

\noindent {\bf Production and Spectroscopy of Baryon-Anti\-baryon Pairs}

Surprisingly little has been measured of the excited (multi)strange and charm 
baryon spectra. This together with the large cross section into baryon-
antibaryon final states (e.g. $\sim 1\ \mu b$ for $\Xi \overline{\Xi}$ or 
$0.1\ \mu b$ for $\Omega \overline{\Omega}$) make spectroscopic studies of 
excited hyperons a very compelling part of the initial program of PANDA when 
the luminosity has not yet reached the design value.
\par
In addition to the hyperon spectra, also the hyperon-hyperon interactions are 
of significant relevance. For example they are an important input for the
physics of compact stars~ \cite{Scha08}.  Traditionally, the core of neutron 
stars has been modeled as a uniform fluid of neutron-rich nuclear matter in 
equilibrium with respect to the weak interaction. Nevertheless, due to the 
large value of the density, other hadronic degrees of freedom ({\it i.e. } 
hyperons) are expected to appear in addition to nucleons \cite{Vida12}.
Unfortunately, in contrast to the nucleon-nucleon interaction, the 
fundamental two- (and multi-) baryon forces are poorly known. So far, the 
world database on Nucleon-Hyperon interaction comprises only a few 
tens of low-momentum $\Lambda$-N and $\Sigma^\pm$-N scattering events, 
very little data on $\Xi$-N and no data on $\Omega $-N scattering 
\cite{Ahn06}. First lattice \cite{Beane:2012ey} and EFT 
\cite{Haidenbauer:2013oca} calculations are emerging, but require better experimental 
{information}. 
%Hypernuclear potentials in dense matter can control the hyperon composition of dense
%neutron star matter. The three-body interactions of nucleons and hyperons can determine the %stiffness of the neutron star equation of state and thereby the maximum neutron star mass. 
%Finally two-body hyperon-nucleon and hyperon-hyperon interactions give rise to hyperon %pairing which exponentially suppresses cooling of neutron stars.
\par
Another example to illustrate the relevance of
hyperon-hyperon interactions is the search for the H-dibaryon. 
The H-dibaryon is an exotic system with strangeness $S = -2$, predicted by Jaffe \cite{Jaffe} and never observed. The discovery of the $_{\Lambda\Lambda}^6$He hypernucleus and the measurement of its binding energy \cite{Taka01} has completely ruled out the possibility that this particle could have a binding energy greater than 7\ MeV. The only open possibility nowadays is a shallow potential or a molecular system.
This state and other strange di-baryons are now amenable to ab initio lattice and also precise
EFT studies, see e.g. \cite{Beane:2010hg,Inoue:2010es,Haidenbauer:2011za}.
At \PANDA{} three different double strange systems will be produced: i) exotic hyperatoms  
(created during the capture process of the $\Xi^-$); ii) doubly strange hypernuclei ($\Xi^-$ 
hypernuclei); and iii) $\Lambda\Lambda$ hypernuclei (following the $\Xi^-$-N interaction). 
We expect to perform $\gamma$-spectroscopy of several tens of double 
$\Lambda$ hypernuclei per month when \PANDA{} is running. This has to be compared 
with less than ten double hypernuclei identified up to now.
\par
Another interesting feature of hyperon study is that they give access to spin observables.  The spin of the hyperon can be related to the spin of the individual quarks. As an example, the spin of the $\Lambda$ hyperon is primarily carried by its valence s quark. Consequently, studying spin variables in the $\bar pp \to \bar YY$ processes probes the role of spin in the creation of strangeness. The same argument holds for the creation of charm in the case of the charmed-hyperons.
%\par

The non-existence of CP violation in strong interactions is an open question. 
The amount of CP violation within the Standard Model is far too low to account for 
the matter-antimatter asymmetry in the universe. It is therefore important to search for 
other sources. CP violation in the hyperon systems can be one of these. Although the 
Standard Model CP violation predictions for hyperons give very low values,  
hyperon CP violation
parameters can be sensitive to effects from physics beyond the Standard
Model such as supersymmetry, left-right symmetric models and multicharged
Higgs \cite{Dono85,Cha95}.
%\vspace{-1mm} 

\section{Summary}

The \PANDA{} experiment will use intense, phase space cooled beams of 
antiprotons to enable precision studies of fundamental questions of hadron 
and nuclear physics in the charm and strangeness sector. Already with the 
start version of FAIR an impressive program of exploration and 
precision measurements will begin, e.g. the search for glueball and 
hybrid candidates, as well as precision width measurements to pin down the 
nature of some of the newly discovered hadronic states. The detector is in an 
advanced stage of preparation and is optimized for high rate resonance and 
threshold scans, with modular components for specific topics such as studies 
of (double) hypernuclei. The completion of the full FAIR facility will provide a factor 20 higher luminosity, enabling the full precision for resolving these fundamental questions.

%The \PANDA{} experiment together with the high quality antiproton beam at the HESR will be a powerful tool to address fundamental questions of hadron physics in the charmed and multi-strange hadron sector. Despite the impressive wealth of recent new observations in the hidden charm meson sector, \PANDA{} will be able to deliver decisive contributions to this field, due to the complementarity of the $\overline{p} p$ entrance channel and the capabilities of the combined storage ring and detector system under construction.

\end{document}